\RequirePackage{ifpdf}
\documentclass[12pt]{JHEP3}         
\usepackage{amssymb,amsfonts}
\usepackage{cite}
\usepackage{amsmath}
\usepackage{graphicx}
\usepackage{MnSymbol}

\newcommand{\p}{\partial}

\newcommand{\ba}{\begin{eqnarray}}
\newcommand{\ea}{\end{eqnarray}}
\newcommand{\be}{\begin{equation}}
\newcommand{\ee}{\end{equation}}
\newcommand{\bal}{\begin{align}}
\newcommand{\eal}{\end{align}}
\newcommand{\bay}[1]{\left(\begin{array}{#1}}
\newcommand{\eay}{\end{array}\right)}

\newcommand{\Tr}{\mbox{Tr}}

\preprint{}
\title{\center Linking Entanglement and Discrete Anomaly}

\author{Ling-Yan Hung${}^{1,2,3}$\ ,~Yong-Shi Wu${}^{3,1,2,4}$\ ,~Yang Zhou${}^1$\\

${}^1$ Department of Physics and Center for Field Theory and Particle Physics, Fudan University, Shanghai 200433, China\\
${}^2$ Collaborative Innovation Center of Advanced Microstructures, Nanjing 210093, China\\
${}^3$ State Key Laboratory of Surface Physics, Fudan University, Shanghai 200433, China\\
${}^4$ Department of Physics and Astronomy, University of Utah, Salt Lake City, Utah, 84112, U.S.A\\

{\tt E-mails : electron.janethung@gmail.com, yswu@fudan.edu.cn, yangzhou1984@gmail.com}
}

\abstract{In $3d$ Chern-Simons theory, there is a discrete one-form symmetry, whose symmetry group is isomorphic to the center of the gauge group. We study the 't Hooft anomaly associated to this discrete one-form symmetry in theories with generic gauge groups, $A,B,C,D$-types. We propose to detect the discrete anomaly by computing the Hopf state entanglement in the subspace spanned by the symmetry generators and develop a systematical way based on the truncated modular S matrix. We check our proposal for many examples.}

\begin{document}

\pagestyle{plain} \setcounter{page}{1}
\newcounter{bean}
\baselineskip16pt


\section{Introduction}
The goal of this paper is to establish a relation between discrete global  anomaly and multi-boundary entanglement in topological quantum field theory. The former has been used to classify symmetry-protected topological phases~\cite{Chen:2011pg,Hung:2012nf}. The anomalies associated to discrete symmetries are also very useful to detect phases of QCD in $3d$ and $4d$~\cite{Gaiotto:2017yup,Komargodski:2017keh,Gomis:2017ixy,Gaiotto:2017tne}. It is well known that 't Hooft anomaly is powerful to constrain infrared dynamics. On the other hand quantum entanglement is a natural quantity to characterize the long-range physics. Moreover, both the entanglement and anomaly are pure quantum effects without classical analogies. Therefore one might expect to build up some precise connection between the two. In this paper we move a small initial step along this direction. Namely we have established a concrete relation between the quantum entanglement and the discrete anomaly in certain field theories. In particular the simplicity of topological field theories allows us to focus on the global nature of both the entanglement and the anomaly without worrying about the local interactions.

There has been a lot of interest to study entanglement in topological field theories. The most popular observable is the topological entanglement entropy~\cite{Levin:2006zz,Dong:2008ft,Kitaev:2005dm}, which is associated to partitions in a single boundary of a Euclidean $d$-dimensional manifold where the topological theories are living on. It is an interesting open question to understand the possible patterns of entanglement that can arise in field theory. One might expect that the entanglement encoded in wavefunctions is always a result of interactions through local Hamiltonians. In fact, entanglement is a property encoding physics beyond locality.

Another way to study entanglement in field theories is to think about entanglement between multi-boundaries. The computation becomes relatively simple in topological quantum field theories. In Euclidean path integral formalism, the $d$-dimensional bulk path integral can be used to prepare the entanglement between different disconnected regions for the boundary state.

The multi-boundary entanglement was first studied in AdS$_3$/CFT$_2$~\cite{Balasubramanian:2014hda,Marolf:2015vma}, later in $3d$ Chern-Simons theory where the entanglement structure of a link state was nicely connected to the framing-independent link invariants~\cite{Salton:2016qpp,Balasubramanian:2016sro}. Recently the study of entanglement structure was further carried out in Chern-Simons theories with generic gauge groups~\cite{Dwivedi:2017rnj}. It is therefore desirable to extract more physical information out of the large amount of entanglement data, which is one of the motivations of this paper.

In the works~\cite{Salton:2016qpp,Balasubramanian:2016sro,Dwivedi:2017rnj} mentioned above, the Chern-Simons theory was placed on a 3-manifold with a boundary consisting of $n$ topologically linked tori. The Hilbert space is the $n$-fold tensor product $\mathcal{H}^{\otimes n}$, where $\mathcal{H}$ is the Hilbert space of Chern-Simons theory on a torus. In this paper, instead of studying the wavefunction in the entire $\mathcal{H}$ space consisting of all integrable representations of the affine Lie algebra, we focus on a subspace spanned by the symmetry generators of a discrete 1-form symmetry. We propose that the reduced density matrix on such a subspace captures all the information about the anomaly of this discrete symmetry.

{\it Note added:} While this paper was in completion, \cite{Numasawa:2017crf} appeared, where the invariant boundary state conditions in WZW models coincide with our $3d$ anomaly free conditions.

 \section{Review of Anomaly of Discrete Symmetry}

Anomaly was originally discovered  as the violation of a classical symmetry at the quantum level, namely the ABJ anomaly~\cite{Bell:1969ts,Adler:1969gk}. This can be interpreted as a 't Hooft anomaly.  The famous axial anomaly is an example of a 't Hooft anomaly for $U(1)_{em}\times U(1)_A$~\cite{Kapustin:2014lwa}. For a global symmetry $G$, the obstruction to be promoted to a gauge symmetry is called 't Hooft anomaly~\cite{tHooft:1979rat}. These anomalies are preserved along renormalization group flows. If $G$ is a connected Lie group, the 't Hooft anomaly is tightly constrained by Wess-Zumino consistency conditions~\cite{Wess:1971yu}. Furthermore, the mechanism of anomaly inflow implies that a 't Hooft anomaly in $d$ space-time dimensions must be classified by possible Chern-Simons actions in $d+1$ dimensions. For finite $G$, Wess-Zumino consistency condition does not apply, but we expect that the inflow mechanism still works. The topological actions in $d+1$ dimensions are classified by elements of the group $H^{d+2}(BG, Z)$~\cite{Dijkgraaf:1989pz}. For finite $G$, we instead have $H^{d+1}(BG,U(1))$, which also classifies bosonic SPT phase with global symmetry $G$ in $d+1$ dimensions~\cite{Chen:2011pg,Hung:2012nf}. 

Another way to see the anomaly is from the commutation relation of the generators of $G$ by noting that the symmetry $G$ is only realized projectively~\cite{Kapustin:2014lwa,Vishwanath:2012tq}. We will see from an example in later discussions that the generators $x$ and $y$ of $Z_k\times Z_k$ do not commute. That is, the symmetry of the theory is a central extension of $Z_k\times Z_k$. 

For continuous symmetries, the 't Hooft anomalies are not affected by the RG flows and therefore constrain the infrared dynamics. When the symmetries are spontaneously broken, the anomalies give rise to Wess-Zumino-Witten terms in the low energy effective action. This is usually called 't Hooft anomaly matching. Since there is no massless Goldstone particles for the spontaneous breaking of discrete symmetries, one can only keep track of the remaining unbroken symmetries in the IR and the UV and impose anomaly matching. 

A higher ($q$-form) symmetry is a global symmetry under which the charged operators are of space-time dimension $q$ and the generators have co-dimension $q+1$ in space-time~\cite{Gaiotto:2014kfa}. This can be contrasted with ordinary symmetries, where $q=0$. The properties of ordinary symmetries can be readily generalized to accommodate higher symmetries. In particular, we will focus on the associated 't Hooft anomalies. 

$3d$ Chern-Simons theory is the simplest example where a higher form symmetry has a 't Hooft anomaly. We will closely follow~\cite{Gaiotto:2014kfa} and review in this section the anomalies they studied. 

\subsection{$U(1)_k$}
$U(1)_k$ Chern-Simons theory has a one-form $Z_k$ symmetry with the generators given by the Wilson loops
\begin{equation}
U_g[W]=\exp\left(in\oint_W A\right)\ , \quad g=e^{i2\pi n/k}\ ,
\end{equation} with $n=0,1,\dots, k-1$. The charged operators are the same Wilson loops with the acting rule
\begin{equation}\label{symmetrydef}
U_g[W]\,U_{g'}[V] = e^{2\pi i mn\over k} U_{g'}[V]\ ,\quad g=e^{2\pi i n/k}\,,g'=e^{2\pi i m/k}\ ,
\end{equation} where $W$ is a circle around $V$.\footnote{Here we mean the linking number is $1$.} It is clear that the generators are charged under themselves, which was interpreted as the appearance of a't Hooft anomaly associated to $Z_k$.

It is of interest to note that in the construction of SET via field extension say in \cite{Hung:2012nf} that embeds the topological order in a larger topological order, it is required that the set of generators of the global symmetries correspond to Bosonic sectors in the extended topological theory, with trivial mutual statistics.  This readily matches with the criterion for an anomaly free symmetry as described above. 

\subsection{$SU(2)_k$}
Another example is $SU(2)_k$ Chern Simons theory. There is a one-form $Z_2$ symmetry generated by the Wilson loop associated with the $SU(2)$ representation with $j=k/2$. The charged operators are the various Wilson loops and the acting rule is
\begin{equation}
U_g[W] C_{j'}[V]=(-1)^{2{j'}}C_{j'}[V],\quad g=e^{{2\pi i\over k}j}=-1\ ,
\end{equation} with $j'=0,{1\over 2},\dots,{k\over 2}$.
This is consistent with the expectation value of two linked loops in the representations $j$ and $j'$ in $S^3$. It is
\begin{equation}
S_{jj'} = \sqrt{2\over k+2}\sin\left(\pi(2j+1)(2j'+1)\over k+2\right)\ .
\end{equation} For $j={k\over 2}$ one has
\begin{equation}
S_{{k\over 2}j'}=(-1)^{2j'}S_{0j'}\ .
\end{equation} Now let us consider the action of generators on themselves
\begin{equation}
U_g[W]U_j[V]=(-1)^{2j}U_j[V], \quad j={k\over 2}\ .
\end{equation}
For odd $k$ there is a 't Hooft anomaly of $Z_2$ symmetry. 
Notice that this $Z_2$ symmetry coincides with the center of the $SU(2)$ gauge group.

\subsection{$SU(N)_k$}
Similarly one can consider $SU(N)_k$ Chern-Simons theory. This theory has a $Z_N$ one-form symmetry. The generator $U_1$ is a Wilson loop with $k$ boxes in a symmetric representation, the other group elements $U_n$ are labeled by rectangular Young Tableaux with $nk$ boxes. Again focus on the acting rule of the generators on themselves
\begin{equation}
U_1[W]U_1[V] = e^{-2\pi i k\over N}U_1[V]\ , 
\end{equation} which indicates that for general $k$ there is a 't Hooft anomaly except that $k$ is a multiple of $N$.

Our observation is that the above anomaly can be directly seen from the linking entanglement, which we will illustrate below.

\section{Entanglement as measure of anomaly}
Consider a 2-component link (such as Hopf link) inside $S^3$ in Chern-Simons theory with gauge group $G$ at level $k$, cut along a tube neighborhood of the link, then there are {\it inside} part and {\it outside} part, the {\it solid torus} and {\it link complement}.
The {\it inside} path integral over solid torus with a Wilson loop insertion, which runs over all integrable representations of affine algebra $G_k$, provides a base of a Hilbert space. Since there are two {\it inside} parts, the total Hilbert space is naturally given by the tensor product of the two. The wave function of the state defined by the {\it outside} path integral in the basis just constructed is precisely the usual link invariant~\cite{Witten:1988hf}. By tracing out one of the sub Hilbert spaces, one can compute the entanglement encoded in this wavefunction (so called multi-boundary entanglement). This is what we call linking entanglement for short. In the following subsection we briefly review the set up in~\cite{Balasubramanian:2016sro}. All the other parts of this section are devoted to the study of the precise relation between linking entanglement and discrete anomaly in different theories.
\subsection{Linking entanglement}
Consider Chern-Simons theory with gauge group $G$ at level $k$, with action defined on 3-manifold
\be
S_{CS}[A] = {k\over 4\pi} \int_M \Tr \left(A\wedge dA + {2\over 3} A\wedge A\wedge A\right)\ .
\ee Consider states defined on $n$ copies of $T^2$, $\Sigma_n$. Then the Hilbert space is the $n$-fold tensor product $\mathcal{H}^{\otimes n}$ as mentioned in the introduction. We construct states by performing the Euclidean path integral of the theory on $M$ with boundary $\p M=\Sigma_n$. There could be many options for $M$ to satisfy this boundary condition, but we choose the link complement of a $n$-component link in $S^3$ as our $M$ following~\cite{Balasubramanian:2016sro}. Then the path integral will produce a state
\be
|L\rangle \in \mathcal{H}^{\otimes n}\ .
\ee
To be specific, let us assign a basis for the Hilbert space $\mathcal{H}$ and write
\be\label{outstate}
|L\rangle = \sum_{j_1,\dots,j_n} C_{L}(j_1,\dots,j_n) | j_1,\dots,j_n\rangle\ ,
\ee where $\{|j\rangle\}$ can be prepared as the path integral of a solid torus with a Wilson line in the integrable representations $R_j$ inserted along the non-contractible circle. As such, one can write the coefficient in (\ref{outstate})
\be
C_{L}(j_1,\dots,j_n) = \langle j_1,\dots,j_n|L\rangle\ .
\ee This actually means that if we assign in $S^3$ the $n$-component link with Wilson lines in the conjugate representation $R^*_{j_i}$ for the $i$-th component, the coefficient in (\ref{outstate}) is nothing but the expectation value of $n$ Wilson lines
\be
C_{L}(j_1,\dots,j_n) = \left\langle W_{R^*_{j_1}}(L_1)\dots W_{R^*_{j_n}}(L_n)\right\rangle_{S^3}\ .
\ee
By the above construction, we assign a quantum entanglement structure to a link in three-sphere. With the wavefunction $C_{L}(j_1,\dots,j_n)$ at hand and also the total Hilbert space factorized, one can explore the entanglement structure sufficiently.

For instance, one can bi-partition a link $L$ as $L=L^m|L^{n-m}$. The reduced density matrix can be evaluated
\be
\rho = {1\over \langle L|L\rangle}\Tr_{L_1,\dots,L_m}|L\rangle\langle L|\ .
\ee And the entanglement entropy is given by
\be
S = -\Tr_{L_{m+1},\dots,L_n}\left(\rho\ln\rho\right)\ .
\ee
In most cases below, we will focus on Hopf link, where the wave function of the link state is given by the modular S matrix
\be
C_{\text{Hopf}} (j_1,j_2)= S_{j_1,j_2}\ .
\ee
\subsection{$U(1)_k$}
We consider a 2-component link with linking number $n$, for $U(1)_k$ Chern-Simons, the entanglement entropy constructed as above is given by~\cite{Balasubramanian:2016sro}
\begin{equation}\label{AbEE}
S_{EE} = \ln\left({k\over \mathrm{gcd}(k,n)}\right)\ .
\end{equation}
In the case of Hopf link $n=1$, the entanglement entropy is $\ln k$. We interpret the non-vanishing of $S_{EE}$ as the signal of the 't Hooft anomaly of $Z_k$, given that the symmetry actions on operators is defined as (\ref{symmetrydef}). Notice that in the Abelian case the generators of the $Z_k$ symmetry span the whole Hilbert space. One can generalize the symmetry definition (\ref{symmetrydef}) to the case with linking number $n$ between $W$ and $V$, then the faithful symmetry becomes $Z_{k/\mathrm{gcd}(k,n)}$. The rank of the group  precisely matches with the entanglement entropy (\ref{AbEE}).
\subsection{$SU(2)_k$}
In the case of $SU(2)_k$ Chern-Simons theory, the generators of $Z_2$ (Wilson loops) are the spin $0$ and spin ${k\over2}$ representations. We will first review the computation of the entanglement entropy for the Hopf link state.

As mentioned before the state defined by the path integral on a 3-manifold with linked torus boundaries has the wave function
\begin{equation}
S_{j_1j_2}(k)=\sqrt{2\over k+2} \sin\left({\pi(2j_1+1)(2j_2+1)\over k+2}\right)\ ,
\end{equation} where $j_1$ and $j_2$ span $0,{1\over 2},\dots,{k\over 2}$. The reduced density matrix after integrating out $\mathcal{H}_2$ is
\begin{equation}
\rho_{j_1j'_1}(k) = {SS^\dagger\over \mathrm{tr} SS^\dagger} = {1\over k+1}\mathrm{diag}(1,1,\dots,1)\ ,
\end{equation} where ${1\over k+1}$ is the normalization factor since the dimension is $k+1$. To gain information of the anomaly associated to the $Z_2$ symmetry, we instead consider a $2\times 2$ small $D(k)$ matrix constructed from $S_{j_1j_2}(k)$ with $j_1,j_2$ span $0,{k\over 2}$:
\begin{equation}
D_{11}(k)=S_{00}=\sqrt{2\over k+2}\sin\left(\pi\over k+2\right),\quad D_{12}(k)=S_{0{k\over 2}}=\sqrt{2\over k+2}\sin\left({\pi(k+1)\over k+2}\right),
\end{equation}
\begin{equation}
D_{21}(k)=S_{{k\over 2}0}=\sqrt{2\over k+2}\sin\left({\pi(k+1)\over k+2}\right),\quad D_{22}(k)=S_{{k\over 2}{k\over 2}}=\sqrt{2\over k+2}\sin\left({\pi(k+1)^2\over k+2}\right).
\end{equation}
The reduced density matrix after tracing out $\widetilde{\mathcal{H}}_2$ is~\footnote{$\widetilde{\mathcal{H}}_2$ is the sub Hilbert space which spans representations $0,{k\over 2}$.}
\begin{equation}
\rho_{2\times 2}(k)={DD^\dagger\over \mathrm{tr} DD^\dagger}\ .
\end{equation} The von Neumann entropy of $\rho_{2\times 2}(k)$ is
\begin{equation}
S =-\mathrm{tr}\left(\rho_{2\times 2}\ln\rho_{2\times 2}\right)= \mathrm{log}\,2\ ,
\end{equation} only if $k$ is odd and otherwise vanishes. We illustrate $D(100)$ and $D(101)$ as follows
\begin{equation}
D(100) = \left(
\begin{array}{cc}
 \frac{\sin \left(\frac{\pi }{102}\right)}{\sqrt{51}} & \frac{\sin \left(\frac{\pi }{102}\right)}{\sqrt{51}} \\
 \frac{\sin \left(\frac{\pi }{102}\right)}{\sqrt{51}} & \frac{\sin \left(\frac{\pi }{102}\right)}{\sqrt{51}} \\
\end{array}
\right);\quad D(101) = \left(
\begin{array}{cc}
 \sqrt{\frac{2}{103}} \sin \left(\frac{\pi }{103}\right) & \sqrt{\frac{2}{103}} \sin \left(\frac{\pi }{103}\right) \\
 \sqrt{\frac{2}{103}} \sin \left(\frac{\pi }{103}\right) & -\sqrt{\frac{2}{103}} \sin \left(\frac{\pi }{103}\right) \\
\end{array}
\right)\ .
\end{equation}

In the following we will consider more general CS theories of A,B,C and D types.

\subsection{$A_{N-1, N\geq2}$}
We will first illustrate the process to find the anomaly free condition for $SU(N)_k$ Chern-Simons theory. $SU(N)_k$ Chern-Simons theory has a $Z_N$ global symmetry. The integrable representations are labeled by a set of non-negative integers, $\hat a = [a_0, a_1,\dots,a_{N-1}]$, where $(a_1,\dots,a_{N-1})$ are Dynkin labels. The integrability condition reads
\begin{equation}
\phi_1 a_1 +\dots+ \phi_{N-1} a_{N-1} \leq k\ ,
\end{equation} where $(\phi_1,\dots,\phi_{N-1})=(1,\dots,1)$ are the comarks of $su(N)$ algebra and $k$ is the Chern-Simons level. Suppose $R_1$ and $R_2$ are two representations of $SU(N)$. The element $S_{R_1,R_2}$ of the modular S matrix is given by
\begin{equation}
S_{R_1,R_2} = (-i)^{N(N-1)\over 2}{N^{-{1\over 2}}\over (k+N)^{N-1\over 2}} \det \left[M_{R_1,R_2}\right]\ ,
\end{equation} where the matrix $M_{R_1,R_2}$ is a $N\times N$ matrix whose element is defined as
\begin{equation}
M_{R_1,R_2}[i,j]= \exp \left[{2i\pi \phi_{R_1}[i]\phi_{R_2}[j]\over k+N}\right]\ . 
\end{equation} The indexed function $\phi$ has a total of $N$ components and its $i$-th component for a representation $R$ is defined as
\begin{equation}
\phi_R[i]=\ell[i] - i -{\ell\over N}+{N+1\over 2}\ ,\quad i=1,\dots,N\ ,
\end{equation} where $\ell[i]$ is the number of boxes in the $i$-th row of the reduced Young diagram of representation $R$ and $\ell$ is the total number of boxes ($\ell[N]=0$).

One can take the reduction of the large modular S matrix to a small $N\times N$ matrix, where the row and column of the small matrix $D_{N\times N}$ spans the representations generating the $Z_N$ symmetry. Then one can evaluate the von Neumann entropy of a reduced density matrix constructed from the small $D$ matrix and find that
\begin{equation}
S = -\mathrm{tr}\left(\rho_{N\times N}\ln\rho_{N\times N}\right) = 0\ ,
\end{equation} only if $k$ is a multiple of $N$ and otherwise finite. This means that only when $k$ is a multiple of $N$, the theory is anomaly free. Here we should emphasize that the truncation method here is only needed for $SU(N\geq 2)_k$ theory and the $U(1)_k$ theory as our first example is beyond this pattern. There we do not need to truncate a modular matrix to diagnose the anomaly because the symmetry $Z_k$ spans the whole Hilbert space. We illustrate some examples of $D_{N\times N}(k)$ and $\rho_{N\times N}(k)$ as follows

\begin{equation}\label{example1}
D_{4\times 4}(k=2) = {1\over 2\sqrt{6}}\left(
\begin{array}{cccc}
 1 & 1 & 1 & 1 \\
 1 & -1 & 1 & -1 \\
 1 & 1 & 1 & 1 \\
 1 & -1 & 1 & -1 \\
\end{array}
\right)\ ,\quad \rho_{4\times 4} (k=2) = \left(
\begin{array}{cccc}
 \frac{1}{4} & 0 & \frac{1}{4} & 0 \\
 0 & \frac{1}{4} & 0 & \frac{1}{4} \\
 \frac{1}{4} & 0 & \frac{1}{4} & 0 \\
 0 & \frac{1}{4} & 0 & \frac{1}{4} \\
\end{array}
\right) ;
\end{equation}

\begin{equation}
D_{4\times 4}(k=3) = \text{c}\times\left(
\begin{array}{cccc}
 1 & 1 & 1 & 1 \\
 1 & i & -1 & -i \\
 1 & -1 & 1 & -1 \\
 1 & -i & -1 & i \\
\end{array}
\right)\ ,\quad \rho_{4\times 4} (k=3) = \left(
\begin{array}{cccc}
 \frac{1}{4} & 0 & 0 & 0 \\
 0 & \frac{1}{4} & 0 & 0 \\
 0 & 0 & \frac{1}{4} & 0 \\
 0 & 0 & 0 & \frac{1}{4} \\
\end{array}
\right) ;
\end{equation}
\begin{equation} 
\rho_{3\times 3}(k=1)=
\left(
\begin{array}{ccc}
 \frac{1}{3} & 0 & 0 \\
 0 & \frac{1}{3} & 0 \\
 0 & 0 & \frac{1}{3} \\
\end{array}
\right);
\quad\rho_{3\times 3}(k=2) = \left(
\begin{array}{ccc}
 \frac{1}{3} & 0 & 0 \\
 0 & \frac{1}{3} & 0 \\
 0 & 0 & \frac{1}{3} \\
\end{array}
\right)\ .
\end{equation}

Notice that the 1-form symmetry group of $SU(N)_k$ Chern-Simons theory coincides with the center of the gauge group. This is because the outer automorphism group of the affine Lie algebra is isomorphic to the center group.\footnote{For an explicit description of outer-automorphism groups of affine Lie algebras and the proof of this isomorphism, see for instance Section 14.2.1 and Section 14.2.3 in the CFT book by Francesco, Mathieu and S\'en\'echal.} Therefore one can generalize the discussion for $A$-type gauge group to other $B,C,D,E$ types. This will give conditions to justify whether there is 't Hooft anomaly for one-form symmetries in other types of theories. Below we illustrate the details for $B_N$, $C_N$ and $D_N$ theories.
\subsection{$B_{N\geq 3}$}
We will illustrate the process to find the anomaly free condition for $SO(2N+1)_k$ Chern-Simons theory. $SO(2N+1)_k$ Chern-Simons theory has a $Z_2$ global symmetry. The integrable representations are labeled by a set of non-negative integers, $\hat a = [a_0, a_1,\dots,a_{N}]$, where $(a_1,\dots,a_{N})$ are Dynkin labels. The integrability condition reads
\begin{equation}
\phi_1a_1 +\phi_2a_2+\dots+ \phi_{N} a_{N} \leq k\ ,
\end{equation} where $(\phi_1,\phi_2,\dots,\phi_{N-1},\phi_N)=(1,2,\dots,2,1)$ are the comarks of $so(2N+1)$ algebra and $k$ is the Chern-Simons level. Suppose $R_1$ and $R_2$ are two representations of $SO(2N+1)$ with Dynkin labels $R_1=[a_1,\dots,a_N]$ and $R_2=[b_1,\dots,b_N]$. The element $S_{R_1,R_2}$ of the modular S matrix is given by
\begin{equation}
S_{R_1,R_2} = (-1)^{N(N-1)\over 2}{2^{N-1}\over (k+2N-1)^{N\over 2}} \det \left[M_{R_1,R_2}\right]\ ,
\end{equation} where the matrix $M_{R_1,R_2}$ is $N\times N$ matrix whose element is defined as
\begin{equation}
M_{R_1,R_2}[i,j]= \sin \left({2\pi \phi_{R_1}[i]\phi_{R_2}[j]\over k+2N-1}\right)\ . 
\end{equation} The indexed function $\phi$ has a total $N$ components and its $i$-th component for a representation $R$ is defined as
\begin{equation}
\phi_R[i]=\ell_i - i +{2N+1\over 2}\ ,\quad i=1,\dots,N\ ,
\end{equation} where $\ell_i$ is determined by Dynkin labels
\begin{equation}
\ell_i = \sum_{n=i}^{N-1} a_n + {a_N\over 2}\ ,\quad \ell_N = {a_N\over 2}\ .
\end{equation}

One can take the reduction of the large modular S matrix to a small $2\times 2$ matrix, where the row and column of the small matrix span the representations generating $Z_2$ symmetry. Then one obtains a reduced density matrix from the small $D$ matrix and find that
\begin{equation}
S=-\mathrm{tr}\left(\rho_{2\times 2}\ln\rho_{2\times 2}\right) =0\ ,
\end{equation} for any $N\geq 3$ and any $k$. This means that the theories are always anomaly free.
\subsection{$C_{N\geq 2}$}
We will illustrate the process to find the anomaly free condition for $Sp(2N)_k$ Chern-Simons theory. $Sp(2N)_k$ Chern-Simons theory has a $Z_2$ global symmetry. The integrable representations are labeled by a set of non-negative integers, $\hat a = [a_0, a_1,\dots,a_N]$, where $(a_1,\dots,a_N)$ are Dynkin labels. The integrability condition reads
\begin{equation}
\phi_1a_1 +\phi_2a_2+\dots+ \phi_{N} a_{N} \leq k\ ,
\end{equation} where $(\phi_1,\phi_2,\dots,\phi_N)=(1,1,\dots,1)$ are the comarks of $sp(2N)$ algebra and $k$ is the Chern-Simons level. Suppose $R_1$ and $R_2$ are two representations of $Sp(2N)$ with Dynkin labels $R_1=[a_1,\dots,a_N]$ and $R_2=[b_1,\dots,b_N]$. The element $S_{R_1,R_2}$ of the modular S matrix is given by
\begin{equation}
S_{R_1,R_2} = (-1)^{N(N-1)\over 2}\left({2\over k+N+1}\right)^{N\over 2} \det \left[M_{R_1,R_2}\right]\ ,
\end{equation} where the matrix $M_{R_1,R_2}$ is $N\times N$ matrix whose element is defined as
\begin{equation}
M_{R_1,R_2}[i,j]= \sin \left({\pi \phi_{R_1}[i]\phi_{R_2}[j]\over k+N+1}\right)\ . 
\end{equation} The indexed function $\phi$ has total $N$ components and its $i$-th component for a representation $R$ is defined as
\begin{equation}
\phi_R[i]=\ell_i - i +N+1\ ,\quad i=1,\dots,N\ ,
\end{equation} where $\ell_i$ is determined by Dynkin labels
\begin{equation}
\ell_i = \sum_{n=i}^N a_n\ .
\end{equation}

One can take the reduction of the large modular S matrix to a small $2\times 2$ matrix, where the row and column of the small matrix span the representations generating the $Z_2$ symmetry. Then one can obtain a reduced density matrix from the small $D$ matrix and find that
\begin{equation}
S=-\mathrm{tr}\left(\rho_{2\times 2}\ln\rho_{2\times 2}\right)> 0\ ,
\end{equation} only if $N$ is odd and also $k$ is odd, but vanishes otherwise. This means that there is a 't Hooft anomaly only for the case where both $N$ and $k$ are odd.
\subsection{$D_{N\geq 4}$}
We will illustrate the process to find the anomaly free condition for $SO(2N)_k$ Chern-Simons theory. $SO(2N)_k$ Chern-Simons theory has a $Z_4$ global symmetry when $N$ is odd and a $Z_2\times Z_2$ global symmetry when $N$ is even. The integrable representations are labeled by a set of non-negative integers, $\hat a = [a_0, a_1,\dots,a_N]$, where $(a_1,\dots,a_N)$ are Dynkin labels. The integrability condition reads
\begin{equation}
\phi_1a_1 +\phi_2a_2+\dots+ \phi_{N} a_{N} \leq k\ ,
\end{equation} where $(\phi_1,\phi_2,\dots,\phi_{N-2},\phi_{N-1},\phi_N)=(1,2,\dots,2,1,1)$ are the comarks of $so(2N)$ algebra and $k$ is the Chern-Simons level. Suppose $R_1$ and $R_2$ are two representations of $SO(2N)$ with Dynkin labels $R_1=[a_1,\dots,a_N]$ and $R_2=[b_1,\dots,b_N]$. The element $S_{R_1,R_2}$ of modular S matrix is given by
\begin{equation}
S_{R_1,R_2} = (-1)^{N(N-1)\over 2}{2^{N-2}\over (k+2N-2)^{N\over 2}} \left(\det \left[M_{R_1,R_2}\right]+i^N\det\left[G_{R_1,R_2}\right]\right)\ ,
\end{equation} where the matrix $M_{R_1,R_2}$ and $G_{R_1,R_2}$ are $N\times N$ matrix whose element is defined as
\begin{equation}
M_{R_1,R_2}[i,j]= \cos \left({2\pi \phi_{R_1}[i]\phi_{R_2}[j]\over k+2N-2}\right)\ ,\quad G_{R_1,R_2}[i,j]=\sin \left({2\pi \phi_{R_1}[i]\phi_{R_2}[j]\over k+2N-2}\right)\ .
\end{equation} The indexed function $\phi$ has total $N$ components and its $i$-th component for a representation $R$ is defined as
\begin{equation}
\phi_R[i]=\ell_i - i +N\ ,\quad i=1,\dots,N\ ,
\end{equation} where $\ell_i$ is determined by Dynkin labels
\begin{equation}
\ell_i = \sum_{n=i}^{N-2} a_n + {a_N+a_{N-1}\over 2}\ ,\quad \ell_{N-1} = {a_N+a_{N-1}\over 2}\ ,\quad\ell_N = {a_N-a_{N-1}\over 2}\ .
\end{equation}

One can take the reduction of the large modular S matrix to a small $4\times 4$ matrix, where the row and column of the small matrix span the representations generating $Z_4$ or $Z_2\times Z_2$ symmetry. Then one can obtain a reduced density matrix from the small $D$ matrix and find that
\begin{equation}
S=-\mathrm{tr}\left(\rho_{4\times 4}\ln\rho_{4\times 4}\right)= 0\ ,
\end{equation} only if $k$ is even when $N$ is even or $k=4\mathbb{Z}$ when $N$ is odd. This means that the theories are anomaly free only for these two cases. We illustrate some examples of $D_{4\times 4}(N,k)$ matrices and $\rho_{4\times 4}(N,k)$ matrices as follows

\begin{equation}
D_{4\times 4}(N=5,k=1) = \left(
\begin{array}{cccc}
 \frac{1}{2} & \frac{1}{2} & \frac{1}{2} & \frac{1}{2} \\
 \frac{1}{2} & \frac{i}{2} & -\frac{i}{2} & -\frac{1}{2} \\
 \frac{1}{2} & -\frac{i}{2} & \frac{i}{2} & -\frac{1}{2} \\
 \frac{1}{2} & -\frac{1}{2} & -\frac{1}{2} & \frac{1}{2} \\
\end{array}
\right); \quad D_{4\times 4}(N=5,k=4)={2-\sqrt{3}\over 24}\left(
\begin{array}{cccc}
 1 & 1 & 1 & 1 \\
 1 & 1 & 1 & 1 \\
 1 & 1 & 1 & 1 \\
 1 & 1 & 1 & 1 \\
\end{array}
\right);
\end{equation}
\begin{equation}\label{example2}
D_{4\times 4}(N=5,k=2) ={1\over 2\sqrt{10}}\left(
\begin{array}{cccc}
 1 & 1 & 1 & 1 \\
 1 & -1 & -1 & 1 \\
 1 & -1 & -1 & 1 \\
 1 & 1 & 1 & 1 \\
\end{array}
\right)\ ,\quad \rho_{4\times 4}(N=5,k=2) = \left(
\begin{array}{cccc}
 \frac{1}{4} & 0 & 0 & \frac{1}{4} \\
 0 & \frac{1}{4} & \frac{1}{4} & 0 \\
 0 & \frac{1}{4} & \frac{1}{4} & 0 \\
 \frac{1}{4} & 0 & 0 & \frac{1}{4} \\
\end{array}
\right);
\end{equation}
\begin{equation}
D_{4\times 4}(N=4,k=3)=\frac{1}{9} \left(2 \sin \left(\frac{\pi }{18}\right)+\cos \left(\frac{\pi }{9}\right)-\cos \left(\frac{2 \pi }{9}\right)\right)\times\left(
\begin{array}{cccc}
 1 & 1 & 1 & 1 \\
 1 & 1 & -1 & -1 \\
 1 & -1 & 1 & -1 \\
 1 & -1 & -1 & 1 \\
\end{array}
\right)\ ,
\end{equation}
\begin{equation}
\rho_{4\times 4}(N=4,k=3)=\left(
\begin{array}{cccc}
 \frac{1}{4} & 0 & 0 & 0 \\
 0 & \frac{1}{4} & 0 & 0 \\
 0 & 0 & \frac{1}{4} & 0 \\
 0 & 0 & 0 & \frac{1}{4} \\
\end{array}
\right);\quad D_{4\times 4}(N=4,k=2) ={1\over 4\sqrt{2}}\left(
\begin{array}{cccc}
 1 & 1 & 1 & 1 \\
 1 & 1 & 1 & 1 \\
 1 & 1 & 1 & 1 \\
 1 & 1 & 1 & 1 \\
\end{array}
\right)\ .
\end{equation}
\begin{equation}
D_{4\times 4}(N=6,k=2)={1\over 4\sqrt{3}}\left(
\begin{array}{cccc}
 1 & 1 & 1 & 1 \\
 1 & 1 & 1 & 1 \\
 1 & 1 & 1 & 1 \\
 1 & 1 & 1 & 1 \\
\end{array}
\right)\ ,\quad D_{4\times 4}(N=6,k=4)= c\times \left(
\begin{array}{cccc}
 1 & 1 & 1 & 1 \\
 1 & 1 & 1 & 1 \\
 1 & 1 & 1 & 1 \\
 1 & 1 & 1 & 1 \\
\end{array}
\right)\ .
\end{equation}
\section{Discussion}
Now we discuss what the entropy we have constructed actually counts. As noted in some examples we have discussed, the entropy actually counts how many symmetries are anomalous. More precisely, the entropy is
\be
S = \log \mathcal{D}\ ,
\ee where $\mathcal{D}$ is the rank of the anomalous group. Let us see a few examples. In the case of $SU(4)_{k=2}$ CS theory (\ref{example1}), the one-form symmetry is $Z_4$, but the entropy is $S = \log 2$ and the eigenvalues of the $4\times 4$ reduced density matrix is $({1\over 2},{1\over 2},0,0)$. This is because only a subgroup $Z_2$ of $Z_4$ is anomalous. The same thing happens for $SO(10)_{k=2}$ CS theory (\ref{example2}). One can therefore examine which subgroup of the global symmetry is anomalous by looking at the entropy.

Another physical meaning of $\mathcal{D}$ in $U(1)_k$ theory is the ground state degeneracy. Let us explain by looking at the $U(1)_k$ CS quantized on a torus. We refer to~\cite{Witten:2015aoa} for recent reviews.
The gauge invariant operators are Wilson loops. There are two different cycles so there will be two different Wilson loop operators $W_a$ and $W_b$. From the quantization condition, $W_{a,b}$ must obey the algebra
\begin{equation}\label{commurel}
W_{a}W_{b}=e^{2\pi i \over k}W_{b}W_{a}\ .
\end{equation}
$W_{a}$ and $W_{b}$ are the generators of a symmetry $Z_k\times Z_k$. This commutation relation allows us to see the anomaly since $Z_k\times Z_k$ is only realized projectively. The actual symmetry is a central extension of $Z_k\times Z_k$. This interpretation is useful to distinguish the anomaly we are discussing from other types of anomaly (such as mixed anomalies).

One can immediately see that (\ref{commurel}) can not be realized on a single vacuum~\cite{Wen:1990zza,Sato:2006de}. The minimal representation has dimension $k$ , with the acting rules
\begin{equation}
W_a|j\rangle =e^{{2\pi i\over k}j}|j\rangle, \quad W_b|j\rangle = | j+1\rangle\ ,\quad j=0,1,\dots, k-1\ .
\end{equation}
So the ground state degeneracy is $k$, which matches with the entropy we have constructed (\ref{AbEE}), $S=\log k$. It is interesting to ask which degrees of freedom could respect the 't Hooft anomaly (of one-form symmetries) in non-abelian CS theories.

\section*{Acknowledgement}
We are grateful for helpful discussions with Siddharth Dwivedi and Adar Sharon.

\appendix

\end{document}